\documentclass[runningheads]{llncs}
\usepackage{graphicx}
\usepackage{tikz}
\usepackage{comment}
\usepackage{amsmath,amssymb} % define this before the line numbering.
\usepackage{color}

\usepackage{microtype}
\usepackage{graphicx}
\usepackage{subcaption}
\usepackage{booktabs} % for professional tables
\usepackage{algorithm,algpseudocode}
\usepackage{amsmath,amssymb,enumitem}
\usepackage{hyperref}
\usepackage{xcolor}
 \makeatletter
 \newcommand{\printfnsymbol}[1]{%
   \textsuperscript{\@fnsymbol{#1}}%
 }
 \makeatother

\usepackage[font=small,labelfont=bf]{caption}
\begin{document}
% \linenumbers
\pagestyle{headings}
\mainmatter
\def\ECCVSubNumber{7627}  % Insert your submission number here

\title{Solving Phase Retrieval with a Learned Reference} % Replace with your title

\begin{comment}
% \titlerunning{ECCV-20 submission ID \ECCVSubNumber} 
% \authorrunning{ECCV-20 submission ID \ECCVSubNumber} 
% \author{Anonymous ECCV submission}
% \institute{Paper ID \ECCVSubNumber}
\end{comment}
%******************

% \begin{comment}
\titlerunning{Solving Phase Retrieval with a Learned Reference}

\author{Rakib Hyder\printfnsymbol{1}
\and
Zikui Cai\thanks{equal contribution}
\and
\index{Asif, M. Salman}M. Salman Asif
}
\authorrunning{R. Hyder, Z. Cai, M. Asif}

\institute{University of California, Riverside, CA 92521, USA \\
\email{\{rhyde001,zcai032,sasif\}@ucr.edu}}
% \institute{University of California Riverside
% \and 

% \end{comment}
%******************
\maketitle

\begin{abstract}
Fourier phase retrieval is a classical problem that deals with the recovery of an image from the amplitude measurements of its Fourier coefficients. Conventional methods solve this problem via iterative (alternating) minimization by leveraging some prior knowledge about the structure of the unknown image. The inherent ambiguities about shift and flip in the Fourier measurements make this problem especially difficult; and most of the existing methods use several random restarts with different permutations. In this paper, we assume that a known (learned) reference is added to the signal before capturing the Fourier amplitude measurements. Our method is inspired by the principle of adding a reference signal in holography. To recover the signal, we implement an iterative phase retrieval method as an unrolled network. Then we use back propagation to learn the reference that provides us the best reconstruction for a fixed number of phase retrieval iterations. We performed a number of simulations on a variety of datasets under different conditions and found that our proposed method for phase retrieval via unrolled network and learned reference provides near-perfect recovery at fixed (small) computational cost. We compared our method with standard Fourier phase retrieval methods and observed significant performance enhancement using the learned reference.\let\thefootnote\relax\footnotetext{Our code is available at \url{https://github.com/CSIPlab/learnPR_reference}}

% \keywords{Phase retrieval, reference design, unrolled network, machine learning.}
\end{abstract}

\section{Introduction}
The problem of \emph{phase retrieval} refers to the challenge of recovering a real- or complex-valued signal from its amplitude measurements. This problem arises in diffraction imaging, X-ray crystallography, and ptychography~\cite{fienup1982phase,gerchberg1972practical,harrison1993phase,millane1990phase,shechtman2015phase}.  
Fourier phase retrieval is a special class of phase retrieval problems aimed at the recovery of a signal from the amplitude of its Fourier coefficients. Let us assume that Fourier amplitude measurements are given as 
 \begin{equation}
     y=|Fx|+\eta,
\end{equation}
where $F$ denotes the Fourier transform operator, $x$ denotes the unknown signal or image, and $\eta$ denotes the measurement noise. Our goal is to recover $x$ given $y$. 

Fourier phase retrieval is essential in many applications, especially in optical coherent imaging. Classical methods for phase retrieval utilize the prior knowledge about the support and positivity of the signals \cite{fienup1982phase,gerchberg1972practical}. Subsequent work has considered the case where the unknown signal is \emph{structured} and belongs to a low-dimensional manifold that is known \emph{a priori}. Examples of such low-dimensional structures include sparsity~\cite{wang2016sparse,Jagatap2017}, low-rank~\cite{ptychTCI,ptycholowrankICASSP}, or neural generative models~\cite{phaseICASSP19,jagatap2019algorithmic}. 
Other techniques like Amplitude flow  \cite{wang2017solving} and Wirtinger flow use alternating minimization \cite{candes2015phase}. Many of these newer algorithms involve solving a \emph{non-convex} problem using iterative, gradient-based methods; therefore, they need to be carefully initialized. The initialization technique of choice is spectral initialization, first proposed in the context of phase retrieval in~\cite{netrapalli2013phase}, and extended to the sparse signal case in~\cite{wang2016sparse,Jagatap2017}.

Fourier phase retrieval problem does not satisfy the assumptions needed for successful spectral initialization and remains highly sensitive to the initialization choice. Furthermore, Fourier amplitude measurements have the so-called trivial ambiguities about possible shifts and flips of the images. Therefore, many Fourier phase retrieval methods test a number of random initializations with all possible flips and shifts and select the estimate with the best recovery error \cite{metzler2018prdeep}. 

In this paper, we assume that a known (learned) reference is added to the signal before capturing the Fourier amplitude measurements. The main motivation for this comes from the empirical observation that knowing a part of the image can often help resolve the trivial ambiguities \cite{barmherzig2019holographic,guizar2007holography,hyder2019asilomar}. We extend this concept and assume that a known reference signal is added to the target signal and aim to recover the target signal from the Fourier amplitude of the combined signal. Adding a reference may not feasible in all cases, but our method will be applicable whenever we can add a reference or split the target signal into known and unknown parts. We can describe the Fourier amplitude (phaseless) measurements with a known reference signal $u$ as 
 \begin{equation}\label{eq:pr_u} 
     y=|F(x+u)|+\eta. 
\end{equation}
Similar reference-based measurements and phase retrieval problems also arise in holographic optical coherence imaging \cite{nolte2011optical}.

\begin{figure}[t]
\begin{center}
\centerline{\includegraphics[width=0.95\textwidth]{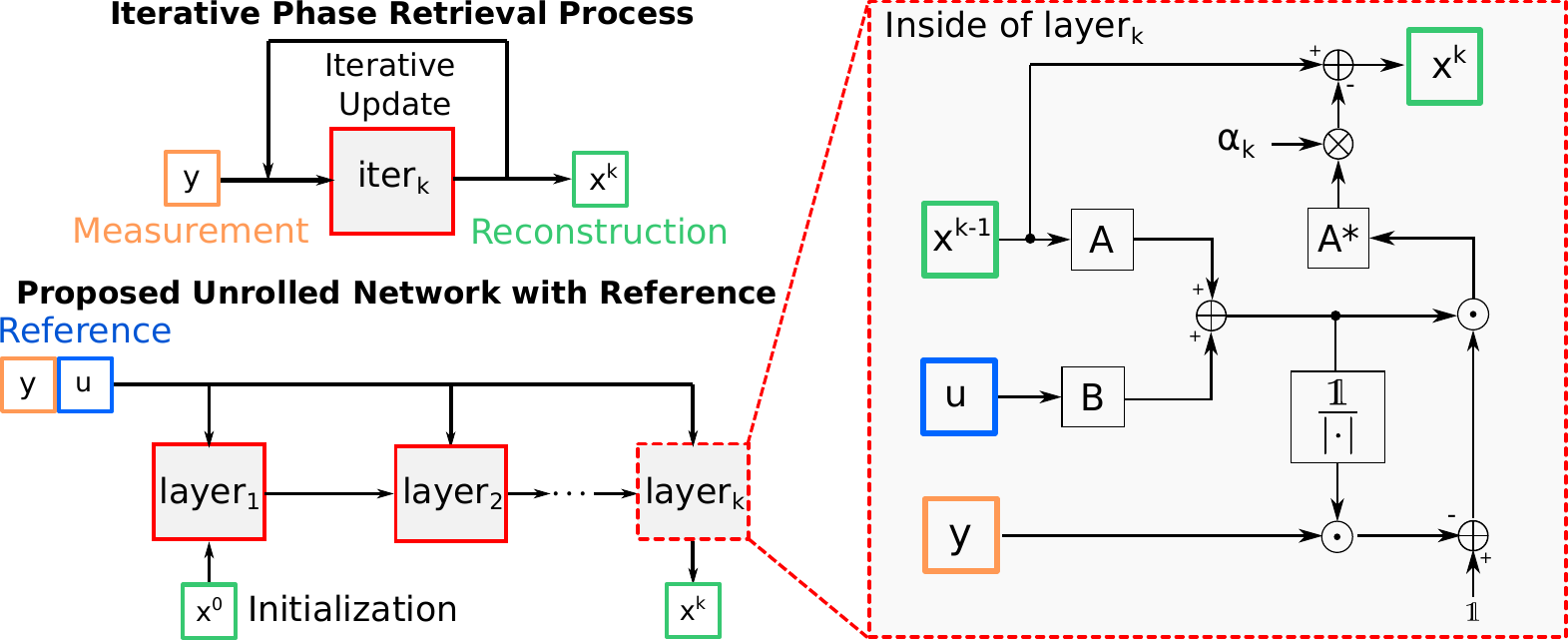}}
\caption{Our proposed approach for learning reference signal by solving phase retrieval using an unrolled network. Unrolled network has $K$ layers. Each layer$_k$ gets amplitude measurements $y$, reference $u$, and estimate $x^{k-1}$ as inputs, and updates the estimate to $x^k$. The operations inside layer$_k$ are shown in the dashed box on the right, where $A$ and $B$ are both linear measurement operators, and $A^*$ is the adjoint operator of $A$. }
\label{fig:intro}
\end{center}
\end{figure}

Our goal is to recover the signal $x$ from the amplitude measurements in \eqref{eq:pr_u}. To do that, we implement a gradient descent method for phase retrieval. We present the algorithm as an unrolled network for a general system in Fig.~\ref{fig:intro}. Every layer of the network implements one step of the gradient descent update. To minimize the computational complexity of the recovery algorithm, we seek to minimize the number of iterations (hence the layers in the network). In addition, we seek to learn the reference $u$ to maximize the accuracy of the recovered signal for a given number of iterations. The learned $u$ and reconstruction results for different datasets are summarized in Fig.~\ref{fig:learnRef_summary}.

\subsection{Our Contributions}
We present an iterative method to efficiently recover a signal from the Fourier amplitude measurements using a fixed number of iterations. To achieve this goal, we first learn a reference signal that can be added to the phaseless Fourier measurements to enable the exact solution of the phase retrieval problem.  
We demonstrate that the reference learned on a very small training set perform remarkably well on the test dataset. 

% Since our approach can reconstruct the signal in a fixed number of steps. This enables efficient applications where real-time is an issue. 
% We also provide a comprehensive theoretical analysis on the one-step reconstruction.

Our main contributions can be summarized as follows. 
\begin{itemize}% [leftmargin=10pt]
    \item The proposed method uses a fixed number of gradient descent iterations (i.e., fixed computational cost) to solve the Fourier phase retrieval problem. 
    \item We formulate the gradient descent method as an unrolled network that allows us to learn a robust reference signal for a class of images. We demonstrate that reference learned on a very small dataset performs remarkably well on diverse and large test datasets. To the best of our knowledge, this is the first work on learning a reference for phase retrieval problems.
    \item We tested our method extensively on different challenging datasets and demonstrated the superiority of our method. 
    %
    % \item Our proposed approach can resolve trivial ambiguities in Fourier phase retrieval problem. We showed comparison with several classical phase retrieval algorithms to demonstrate that.
    \item We demonstrate the robustness of our approach by testing it with the noisy measurements using the reference that was trained on noise-free measurements. % Our results indicate that the proposed method is stable under different levels of Gaussian and Poisson noise.
    % $\item We showed cross dataset testing with learned reference. We observed that reference learned on one dataset performs also well on the datasets with similar distribution.
    % \item We provided comprehensive analysis on the one-step reconstruction.
\end{itemize} 

\section{Related Work}

\textbf{Holography.}
 Digital holography is an interferometric imaging technique that does not require the use of any imaging lens. Utilizing the theory of diffraction of light, a hologram can be used to reconstruct three-dimensional (3D) images \cite{park2018characterization}. With this advantage, holography can be used to perform simultaneous imaging of multidimensional information, such as 3D structure, dynamics, quantitative phase, multiple wavelengths, and polarization state of light \cite{tahara2018digital}. 
%  We can find its applications in natural 3D displays and lensless 3D image recording techniques. 
 In the computational imaging community, many attempts have been made in solving holographic phase retrieval using references, among which \cite{barmherzig2019holographic} has been very successful. Motivated by the reference design for holographic phase retrieval, we are trying to explore a way to design references for general phase retrieval.

\textbf{Phase Retrieval.} The phase retrieval problem has drawn considerable attention over the years, as many optical detection devices can only measure amplitudes of the Fourier transform of the underlying object (signal or image). Fourier phase retrieval is a particular instance of this problem that arises in optical coherent imaging, where we seek to recover an image from its Fourier modulus \cite{fienup1982phase,gerchberg1972practical,rodenburg2008ptychography,millane1990phase,shechtman2015phase,maiden2009improved}.
Existing algorithms for solving phase retrieval can be broadly classified into convex and non-convex approaches \cite{hyder2019alternating}. Convex approaches usually solve a constrained optimization problem after lifting the problem. The PhaseLift algorithm \cite{candes2013phaselift} and its variations \cite{gross2017improved}, \cite{candes2015phasediff} belong to this class. On the other hand, non-convex approaches usually depend on Amplitude flow \cite{wang2016sparse,wang2016solving} and Wirtinger flow \cite{candes2015phase,zhang2016reshaped,chen2015solving,cai2016optimal}. If we know some structure of the signal a priori, it helps in the reconstruction. Sparsity is a very popular signal prior. Some of the approaches for sparse phase retrieval include \cite{ohlsson2012cprl,li2013sparse,bahmani2015efficient,jaganathan2012recovery,netrapalli2013phase,cai2016optimal,wang2016sparse}. Furthermore,  \cite{netrapalli2013phase,Jagatap2017,hyder2019alternating} used  minimization (AltMin)-based approach and \cite{chang2016phase} used total variation regularization to solve phase retrieval. Recently, various researchers have explored the idea of replacing the sparsity priors with generative priors for solving inverse problems. Some of the generative prior-based approaches can be found in \cite{hyder2019alternating,jagatap2019algorithmic,hand2018phase,shamshad2018robust}.

\textbf{Data-Driven Approaches for Phase Retrieval.} 
The use of deep learning-based methods to solve computational imaging problems such as phase retrieval is becoming popular. Deep learning methods leverage the power of huge amounts of data and tend to provide superior performance compared to traditional methods while also run significantly faster with the acceleration of GPU devices. A few examples demonstrating the benefit of the data-driven approaches include \cite{metzler2018prdeep} for robust phase retrieval, \cite{kellman2019data} for Fourier ptychographic microscopy, and \cite{rivenson2018phase} for holographic image reconstruction.

\textbf{Unrolled Network for Inverse Problem.}
Unrolled networks, which are constructed by unrolled iterations of a generic non-linear reconstruction algorithm, have also been gaining popularity for solving inverse problems in recent years \cite{kellman2019physics,diamond2017unrolled,gregor2010learning,wang2016proximal,hammernik2018learning,yang2016deep,kamilov2016learning,bostan2018learning}. Iterative methods usually terminate the iteration when the condition satisfies theoretical convergence properties, thus rendering the number of iterations uncertain. An unrolled network has a fixed number of iterations (and cost) by construction and they produce good results in a small number of steps while enabling efficient usage of training data. 

\textbf{Reference Design.} 
Fourier phase retrieval faces different trivial ambiguities because of the structure of Fourier transformation. As a phase shift in the Fourier domain results in a circular shift in the spatial domain, we will get the same Fourier amplitude measurements for any circular shift of the original signal. In recent papers \cite{barmherzig2019holographic,Yuan_2019,guizar2007holography,hyder2019asilomar}, authors tried to use side information with sparsity prior to mitigate these ambiguities. However, in those studies, the reference and target signal are separated by some margin. 
If the separation between target and reference is large enough, then the nonlinear PR problem simplifies to a linear inverse problem \cite{arab2020fourier,barmherzig2019holographic}. 
%
% Such setup has its own applications in holographic phase retrieval.

In this paper, we consider the reference signal to be additive and overlapping with the target signal. To the best of our knowledge, there has not been any study on such unrestricted reference design. 
While driven by data, our approach for reference design uses training samples in a very efficient way. The number of training images required by our network is parsimonious without limiting its generalizability. The reference learned by our network provides robust recovery test images with different sizes. Apart from the great flexibility, our unrolled network uses a well-defined routine in each layer and demonstrates excellent interpretability as opposed to black-box deep neural networks.

\section{Proposed Approach}

We use the general formulation for the phase retrieval from amplitude measurements. The formulation can be extended for phase retrieval with squared amplitude measurement as well. In our setup, we model amplitude measurements of a target signal $x$ and a reference signal $u$ as $y= |Ax+Bu|$, where $A$ and $B$ are linear measurement operators. Our goal is to learn a reference signal that provides us the best recovery of the target signal. We formulate this overall task as the following optimization problem:
\begin{equation}
    \underset{\hat{x}(u)}{\text{minimize}}\; \|x-\hat{x}(u)\|_2^2 \;\;\;\; \text{s.t.} \;\;    y = |A\hat{x}(u)+Bu|, \label{eq:learnRef}
\end{equation}
where $\hat x(u)$ denotes the solution of the phase retrieval problem for a given reference $u$. 
Our approach to learn $u$ and solve \eqref{eq:learnRef} can be divided into two nested steps: (1) Outer step updates $u$ to minimize the recovery error for phase retrieval and (2) inner step uses the learned $u$ to recover target images by solving phase retrieval.
%

% \subsection{Learning Reference Signal}
% \textbf{Learning Reference Signal.} 
To solve the (inner step) of phase retrieval problem, we use an unrolled network. Figure~\ref{fig:intro} depicts the structure of our phase retrieval algorithm. In the unrolled phase retrieval network, we have $K$ blocks to represent $K$ iterations of the phase retrieval algorithm. We minimize the following loss to solve the phase retrieval problem:
\begin{equation}
    L_{x} (x,u) = \|y - |Ax+Bu| \|_2^2.
    \label{eq:pr_general}
\end{equation}
Every block of the unrolled phase retrieval network is equivalent to one gradient descent step for  \eqref{eq:pr_general}. For some value of reference estimate, $u$, we can represent the target signal estimate after $k+1^{th}$ block of the unrolled network as
\begin{equation}
    x^{k+1} = x^{k} - \alpha_k \nabla_{x}L_{x}(x^{k},u),
    \label{eq:pr}
\end{equation}
where $\nabla_{x}L_{x}(x^{k},u)$ is the gradient of $L_x$ with respect to $x$ at the given values of $x^{k}, u$. As the loss function in  \eqref{eq:pr_general} is not differentiable, we can redefine it as
\begin{equation}
    L_{x} (x,u) = \|y\odot p - (Ax+Bu) \|_2^2,
    \label{eq:pr_lin}
\end{equation}
where $p=\angle(Ax^k+Bu)=(Ax^k+Bu)/|Ax^k+Bu|$. The expression of gradient can be written as 
\begin{equation}
    \nabla_{x}L_{x} (x^k,u) = 2A^* [p \odot (p^*\odot(Ax^k+Bu)-y)], 
    \label{eq:pr_grad}
\end{equation}
where $A^* $ denotes the adjoint of $A$. After $K$ blocks, we get the estimate of the target signal that we denote as  $\hat{x}(u)=x^{K}$.

In the learning phase, we are given a set of training signals, $\{x_1,x_2,...,x_N\}$, which share the same distribution as our target signals. We initialize $x^0$ and $u^0$ with some initial (feasible) values. First we minimize the following loss with respect to $u$: 
\begin{equation}
    L_u (u) =  \sum_{i=1}^{N}\|x_i- \hat x_{i}\|_2^2=\sum_{i=1}^{N}\|x_i-x_{i}^{K}\|_2^2. 
    \label{eq:lu_org}
\end{equation}
We can rewrite \eqref{eq:lu_org} using the gradient recursion in \eqref{eq:pr} as
\begin{equation}
    L_u (u) =  \sum_{i=1}^{N}\|x_i-x_{i}^0+\sum_{k=0}^{K-1} \alpha_k \nabla_{x}L_{x}(x_i^{k},u)\|_2^2. 
    \label{eq:lu}
\end{equation}
We can then use gradient descent to to minimize $L_u (u)$. We can represent the $j+1^{th}$ iteration of gradient descent step as
\begin{equation}
   u^{j+1} = u^{j} - \beta \nabla_{u}L_{u} (u^{j}).
\end{equation}
The expression for $\nabla_{u}L_{u}(u)$ can be written as
\begin{align}
% \resizebox{.95\hsize}{!}
{\nabla_{u}L_{u}(u)= 2\sum_{i=1}^{N}\left[\sum_{k=0}^{K-1} \alpha_k J_u(x_i^k,u)\right]\left[x_i-x_{i}^0+
\sum_{k=0}^{K-1} \alpha_k \nabla_{x}L_{x}(x_i^{k},u)\right]},
\end{align}
where $J_u(x_i^k,u) =  \nabla_{u}\nabla_{x}L_{x}(x_i^{k},u)$ is a Jacobian matrix with rows and columns of the same size as $u$ and $x$, respectively. 
The measurement vector $y=|Ax+Bu|$ is a function of $u$ during training. Since we model $\hat{x}(u)$ as an unrolled network, we can think of the gradient step as a backpropagation step. To compute $\nabla_{u}L_{u}(u)$,  we backpropagate through the entire unrolled network. At the end of $J^{th}$ outer iteration, we will get our learned reference $\hat{u}=u^J$. 

Once we have learned a reference, $\hat u$, we can use it to capture (phaseless) amplitude measurements as $y = |Ax^*+B\hat{u}|$ for target signal $x^*$. To solve the phase retrieval problem, we perform one forward pass through the unrolled network. Pseudocodes for training and testing are provided in Algorithms~\ref{algo:train},\ref{algo:test}.

In our Fourier phase retrieval experiments $A=B=F$, where $F$ is the Fourier transform operation. To implement similar method for squared amplitude measurements, we can simply replace  $p=\angle(Ax^k+Bu^j)$ with $p=Ax^k+Bu^j$. In all our experiments, we initialized $x^0$ as a zero vector whenever $\hat u \ne 0$.  
We can also add additional constraints on the reference while minimizing the loss function in \eqref{eq:lu}. In our experiments, we used target signals with intensity values in the range $[0,1]$; therefore, we restricted the range of entries in $u$  to $[0,1]$ as well. We discuss other constraints in the experiment section. 
%One such constraint we added is the range constraint. We needed our learned reference to have a comparable range as our target signal. So, we constrained our reference to $[0,1]$ range (range of the target signal) in all the cases unless otherwise mentioned. 
%

\begin{algorithm}[tb]
   \caption{Learning Reference Signal}
   \label{algo:train}
\begin{algorithmic}
   \State {\bfseries Input:} Training signals $\{x_1,x_2,...,x_N\}$, measurement operators, $A$ and $B$.
   
   \State Initialize $\{x_1^0,x_2^0,...,x_N^0\}, u^0$
   \For{$j = 0, 1, ..., J-1$}
        \For{$i=1, 2, ..., N$}
            \State $y_i = |Ax_i^*+Bu^j|$
            \For{$k =  0, 1, ..., K-1$}
    			\State $L_{x} (x_i^k,u^j) = \|y_i - |Ax_i^k + Bu^j| \|_2^2$
    			\State $x^{k+1}_i \leftarrow x^k_i - \alpha_k \nabla_{x}L_{x}(x_i^k,u^j)$ 
            \EndFor
        \EndFor
        \State $L_{u}(u^j)  = \sum_{i=1}^{N}\|x_i^*-x_{i}^0 +  \sum_{k=1}^{K} \alpha_k \nabla_{x}L_{x}(x_i^{k-1},u^j)\|_2^2$
        % \Statex
    	\State $u^{j+1} \leftarrow u^j - \beta \nabla_{u}L_{u}(u^j)$
     \EndFor
     \State {\bfseries Output:} Optimal reference, $\hat{u}=u^J$
    %   \Statex
\end{algorithmic}
\end{algorithm}

\begin{algorithm}[tb]
   \caption{Solving Phase Retrieval via Unrolled Network}
   \label{algo:test}
\begin{algorithmic}
   \State {\bfseries Input:}  Measurements $y$, learned reference $\hat{u}$, measurement operators, $A$ and $B$.

   \State Initialize $x^0$
	\For{$k = 0, 1, ..., K-1$}
        \State $L_{x} (x^k,\hat{u}) = \|y - |Ax^k + B\hat{u}| \|_2^2$ 
    	\State $x^{k+1} \leftarrow x^k - \alpha_k \nabla_{x}L_{x}(x^k,\hat{u})$ 
	\EndFor
   \State {\bfseries Output:} Estimation of target signal $\hat{x}=x^K$

\end{algorithmic}
\end{algorithm}

\section{Experiments}

\begin{figure}[t]
%\vskip 0.2in
\begin{subfigure}[b]{0.5\linewidth}
\begin{center}
\centerline{\includegraphics[trim={4cm 0 3cm 0}, clip, width=\columnwidth]{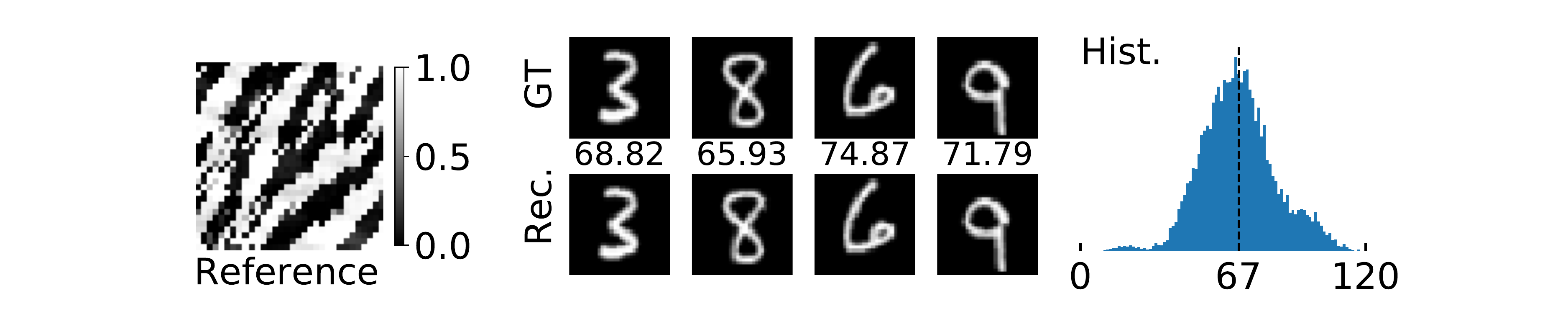}}
\caption{MNIST}
\end{center}
\end{subfigure}
%\vskip -0.2in
%\vskip 0.2in
\begin{subfigure}[b]{0.5\linewidth}
\begin{center}
\centerline{\includegraphics[trim={4cm 0 3cm 0}, clip, width=\columnwidth]{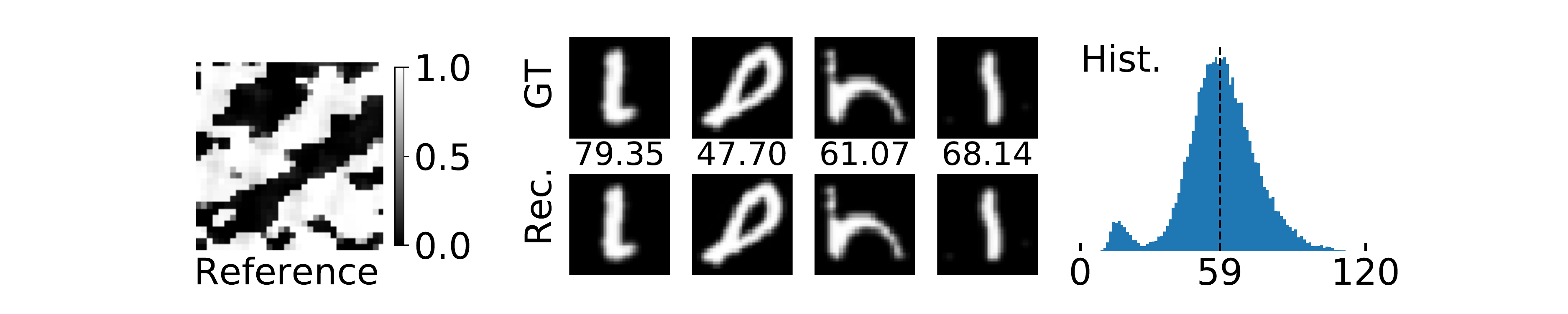}}
\caption{EMNIST}
\end{center}
\end{subfigure}
%\vskip -0.2in
%\vskip 0.2in
\begin{subfigure}[b]{0.5\linewidth}
\begin{center}
\centerline{\includegraphics[trim={4cm 0 3cm 0}, clip, width=\columnwidth]{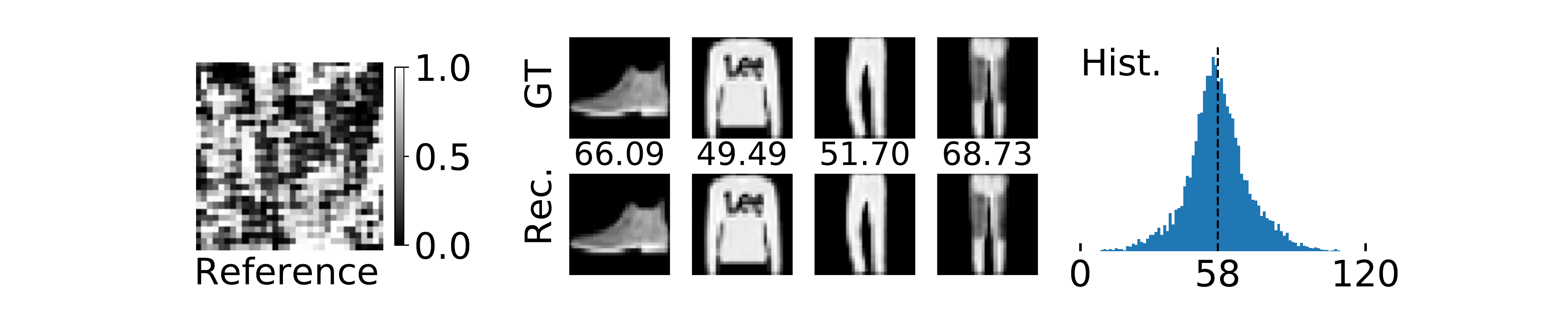}}
\caption{Fashion MNIST}
\end{center}
\end{subfigure}
%\vskip -0.2in
%\vskip 0.2in
\begin{subfigure}[b]{0.5\linewidth}
\begin{center}
\centerline{\includegraphics[trim={4cm 0 3cm 0}, clip, width=\columnwidth]{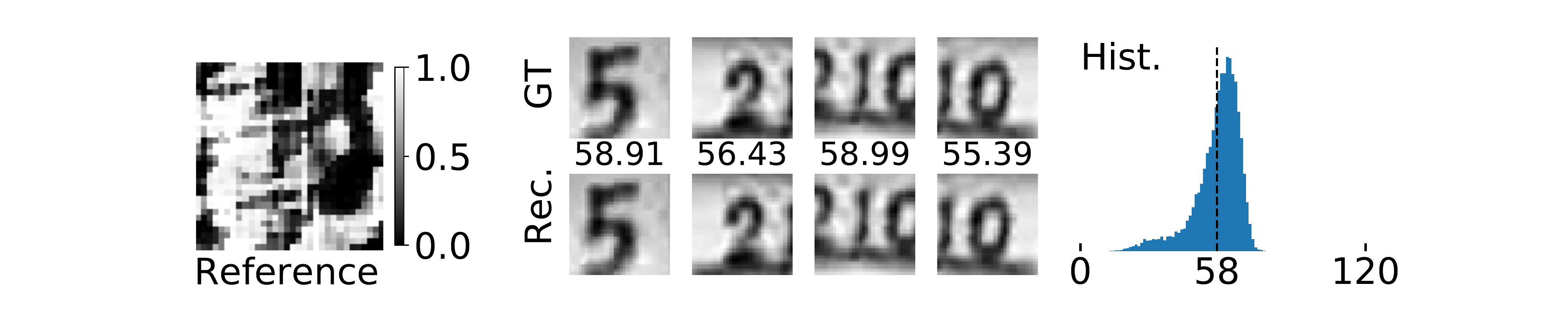}}
\caption{SVHN}
\end{center}
\end{subfigure}
%\vskip -0.2in
%\vskip 0.2in
\begin{subfigure}[b]{0.5\linewidth}
\begin{center}
\centerline{\includegraphics[trim={4cm 0 3cm 0}, clip, width=\columnwidth]{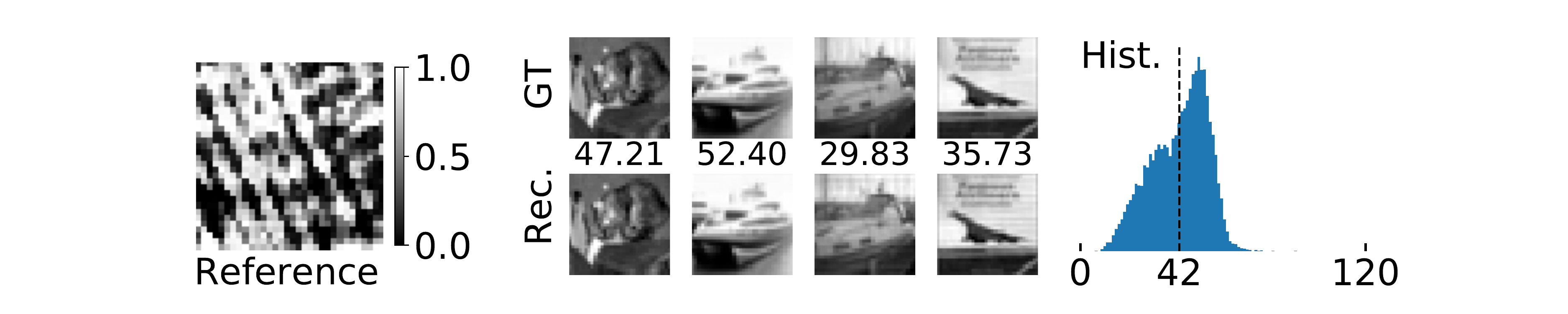}}
\caption{CIFAR10}
\end{center}
\end{subfigure}
%\vskip -0.2in
%\vskip 0.2in
\begin{subfigure}[b]{0.5\linewidth}
\begin{center}
\centerline{\includegraphics[trim={4cm 0 3cm 0}, clip, width=\columnwidth]{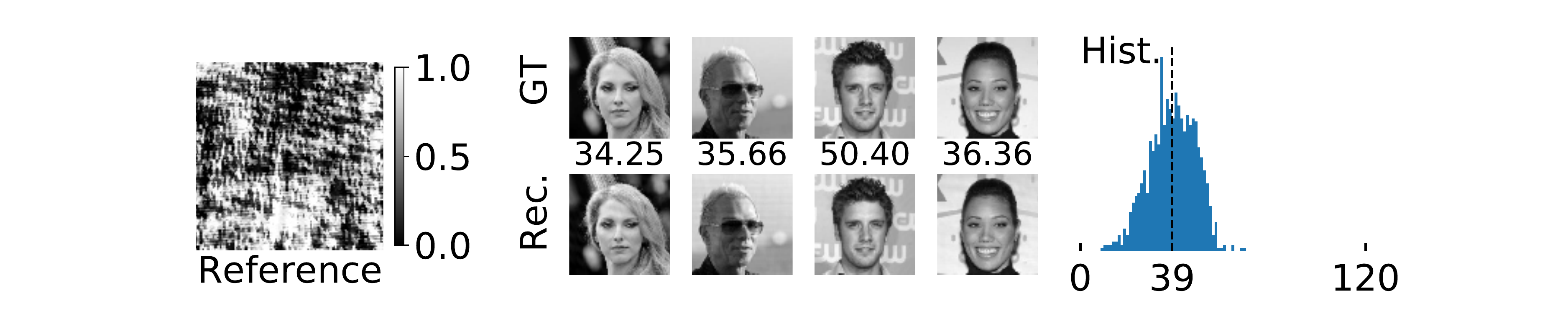}}
\caption{CelebA}
\end{center}
\end{subfigure}
%\vskip -0.2in
\caption{Reconstruction results using learned references. Each block \textbf{(a)-(f)} shows results for a different dataset: (\textbf{left}) learned reference with a colorbar; (\textbf{middle}) sample original images and reconstruction with PSNR on top; (\textbf{right}) histogram of PSNR over the entire test dataset (vertical dashed line represents the mean PSNR).}
\label{fig:learnRef_summary}
\end{figure}

\begin{figure}[t]
\begin{center}
\centerline{\includegraphics[width=1.0\textwidth,trim={0cm 0cm 0cm 0cm},clip]{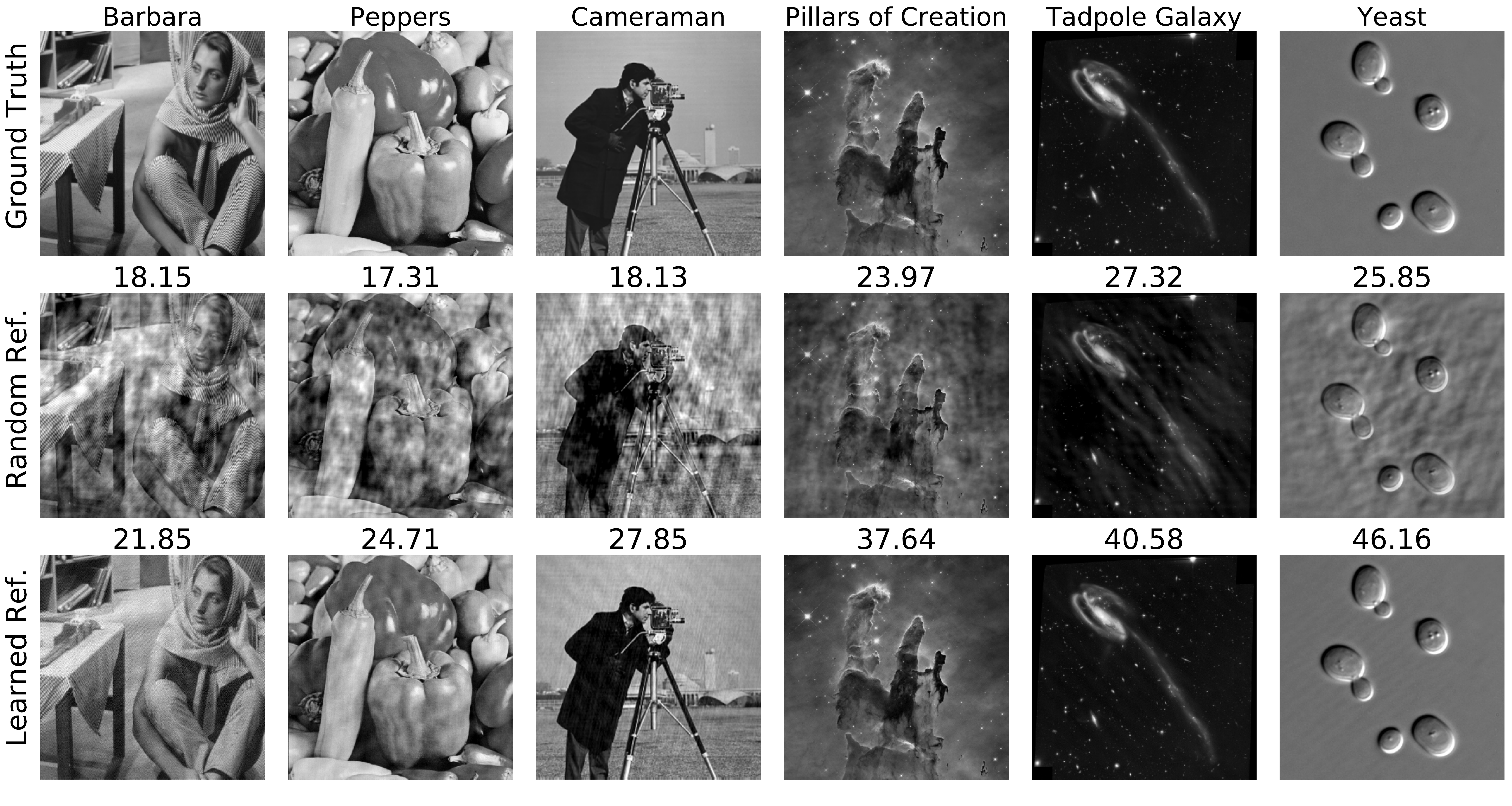}}
\caption{Phase retrieval results using learned and random references. \textbf{First Row:} Original $512 \times 512$ test images. \textbf{Second Row:} Reconstruction using random references with uniform distribution between $[0,1]$ \textbf{best result out of 100 trials}.  \textbf{Third Row:} Reconstruction using the reference learned on CelebA dataset and resized from $200\times 200$ to $512 \times 512$. (PSNR shown on top of images.)
%Our method requires on average 1.24 seconds to recover one image.
}
\label{fig_big_imgs}
\end{center}
\end{figure}

\textbf{Datasets.} 
We have used MNIST digits, EMNIST letters, Fashion MNIST, CIFAR10, SVHN, CelebA datasets, and different well-known standard images for our experiments. We convert all images to grayscale and resize $28\times28$ images to $32\times32$. Although there are tens of thousands training images in MNIST, EMNIST letters, Fashion MNIST, CIFAR10, and SVHN dataset, we have used only a few (e.g.,  32) of them in training. We have shown that the references learned on the small number of training images perform remarkably well on the entire test dataset. MNIST, Fashion MNIST, and CIFAR10 test datasets contain 10000 test images each; EMNIST letters dataset contains 24800 test images; SVHN test dataset contains 26032 test images. We used 1032 images from CelebA and center-cropped and resized all of them to $200\times 200$. We selected 32 images for training and the rest for testing. 

We present the results for these different datasets using references learned from 32 images from the same dataset in Fig.~\ref{fig:learnRef_summary}. We present results for six standard images of size $512\times 512$ from \cite{metzler2018prdeep} using a resized reference learned from CelebA dataset in Fig.~\ref{fig_big_imgs}.

\textbf{Measurements.}
We simulated amplitude measurements of the 2D Fourier transform. We performed 4 times oversampling in the spatial domain for both reference and target signal. Unless otherwise mentioned, we consider our measurements to be noise-free. We also report results for noisy measurements.

\subsection{Configurations of Reference (u)}
 The reference signal $u$, which we are trying to learn, has a number of hyper-parameters that inherently affect the performance of the phase retrieval process. We considered several constraints on $u$, including the support, size, range, position, and sparsity. 
 
 We tested reference signals with both complex and real values and found that $u$ has comparable results in the  two domains. Since it is easy to physically create amplitude or phase-only reference signals, we constrain $u$ to be in the real domain; thus, $u \in \mathbb{R} ^{m \times n}$ and $m$, $n$ represent height and width,  respectively. The height and width of $u$ determine the overlapping area between the target signal and the reference. We found that $u$ with larger size tends to have better performance, especially when the value of $u$ is constrained to a small range. The intensity values of $u$ play a major role in its performance. If we constrain the value of $u$ to be within a certain range: $u[i,j] \in  [u_{min},u_{max}]$, for all $i,j$, we observed that bigger range of $u$ yields better performance. This is because when $u$ is unconstrained then we can construct a $u$ with a large norm. Consider the noiseless setting with quadratic measurements $|F(x+u)|^2 = |Fx|^2 + |Fu|^2 + 2\text{Re}(Fx\odot Fu)$, the last term is the real value of the element-wise product of target and reference Fourier transforms. We can remove $|Fu|^2$ because it is known. If $u$ is large compared to $x$, then we can also ignore the quadratic term $|Fx|^2$ and recover $x$ in a single iteration if all entries of $Fu$ are nonzero. To avoid this situation and make the problem stable in the presence of noise, we restricted the values in the reference $u$ to be in [0,1] range.

\subsection{Setup of Training Samples and Sample Size}
We observed that we can learn the reference signal from a small number of training images. In Table~\ref{table_different_size}, we report test results for different reference signals learned on first $N$ images from MNIST training dataset for $N=32, 128, 512$. We kept the signal and reference strength (i.e., the range of the signal) equal for this experiment. We observe that increasing the training size improves test performance. However, we can get reasonable reconstruction performance on large test datasets (10k+  images) with reference learned using only 32 images.

\begin{table*}[t]
\caption{PSNR for different training sizes}
\label{table_different_size}
\begin{center}
\begin{small}
\begin{sc}
\begin{tabular}{l|cccccc}
\toprule
Train/Test & MNIST & EMNIST & F. MNIST & SVHN  & CIFAR10 \\
\midrule
Training size=32           &   66.54  &  58.72  & 57.81 &  57.51 & 41.60  \\
Training size=128           &   76.25   & 64.16  & 55.86 &  59.50 & 44.34  \\
Training size=512    &   79.14   & 62.34    & 52.01 &  59.78 & 48.90  \\

\bottomrule
\end{tabular}
\end{sc}
\end{small}
\end{center}
\end{table*}

\subsection{Generalization of Reference on Different Classes}
We are interested in evaluating the generalization of our learned reference. (i.e., how the reference performs when trained on one dataset and tested on another). In the comparison study, we took the reference $u$ trained on each dataset and then tested them on the remaining 4 datasets. The value range of the reference is between $[0,1]$, the number of steps in the unrolled network is $K = 50$. We observed that when the datasets share great similarity (e.g., MNIST and EMNIST are both sparse digits or letters), the reference signal tends to work well on both datasets. Even when the datasets differ greatly in their distributions, the reference trained on one dataset provides good results on other datasets (with only a few dB of PSNR decrease in performance). 

\begin{figure}[t]
\centering 
\begin{subfigure}[b]{0.47\linewidth}
\includegraphics[width=1\textwidth]{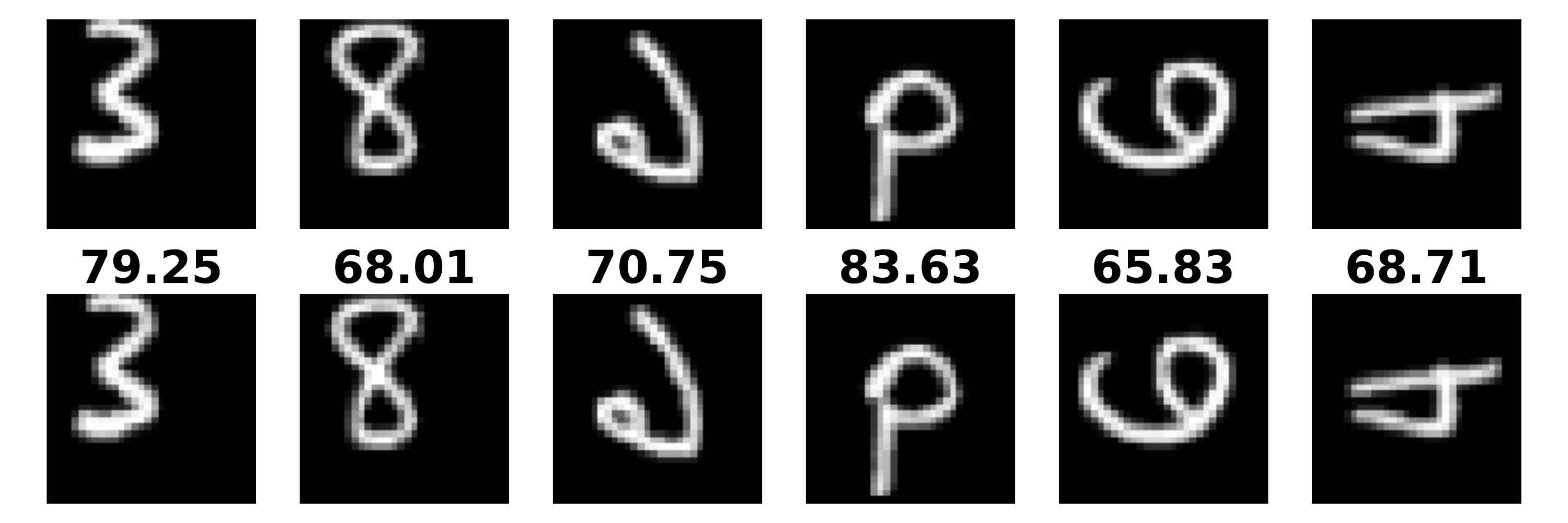}
\caption{MNIST}
\end{subfigure}
~~
\begin{subfigure}[b]{0.47\linewidth}
\includegraphics[width=1\textwidth]{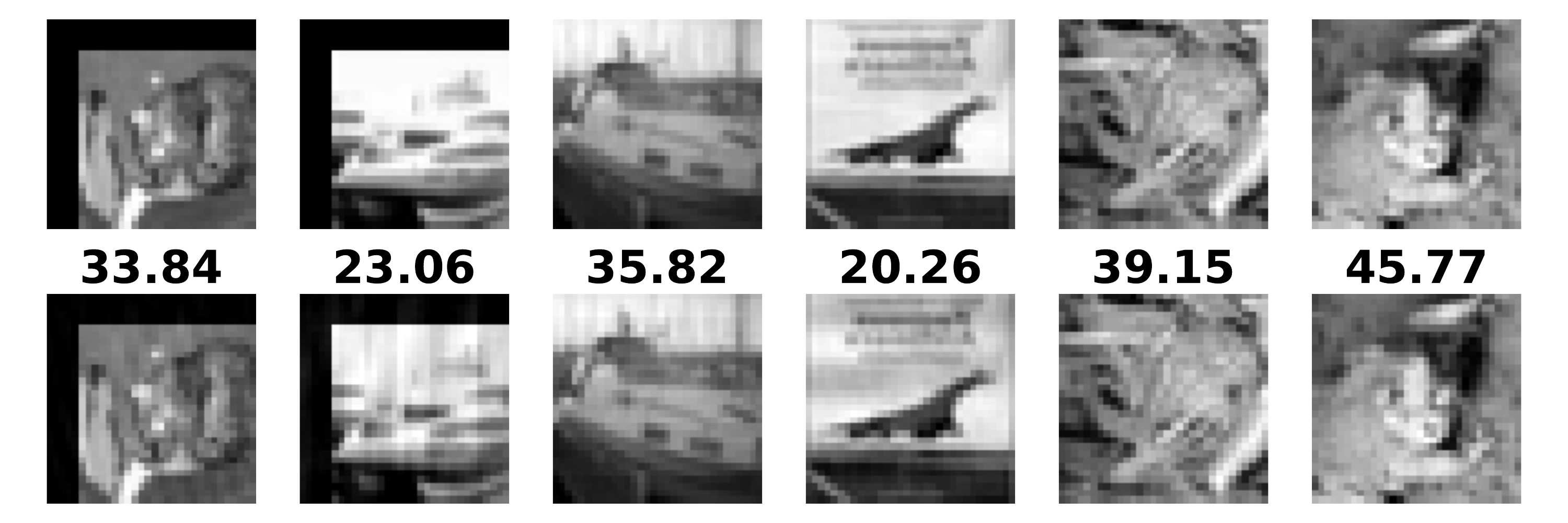}
\caption{CIFAR10 }
\end{subfigure}
\caption{Test results on shifted/flipped/rotated images using the reference learned on upright and centered (canonical) training images. (PSNR shown on top of images.)}
\label{fig:mismatch}
\end{figure}

We also tested our method on shifted and rotated versions of test images. Results in Fig.~\ref{fig:mismatch} demonstrate that even though the reference was trained on upright and centered images, we can perfectly recover shifted and rotated images. 

Our key insight about this generalization phenomenon is that the main challenge in Fourier phase retrieval methods is initialization and ambiguities that arise because of symmetries. We are able to solve these issues using a learned reference because of the following reasons: (1) A reference gives us a good initialization for the phase retrieval iterations. (2) The presence of a reference breaks the symmetries that arise in Fourier amplitude measurements. Moreover, we are not learning to solve the phase retrieval problem in an end-to-end manner or learn a signal-dependent denoiser to solve the inverse problem \cite{metzler2018prdeep,rivenson2018phase}. We are learning reference signals to primarily help a predefined phase retrieval algorithm to recover the true signal from the phaseless measurements. Thus, the references learned on one class of images provide good results on other images, see Table~\ref{table_different_class}. This study shows that the reference learned using our network has the ability to generalize to new datasets, thus making our method suitable for real-life applications where new test cases keep emerging.

\begin{table*}[t]
\caption{PSNR with references trained and tested on different datasets}
\label{table_different_class}
\begin{center}
\begin{small}
\begin{sc}
\begin{tabular}{l|cccccc}
\toprule
Train/Test & MNIST & EMNIST & F. MNIST & SVHN  & CIFAR10 \\
\midrule
MNIST           &   66.54   &  55.12  & 40.87 &  41.87 & 31.72  \\
EMNIST          &   \textbf{72.84}   &  \textbf{58.72}  & 52.18 &  55.42 & 48.16  \\
F. MNIST        &   40.87   &  55.67  & \textbf{57.81} &  50.70 & 42.85  \\
SVHN            &   41.87   &  46.76  & 49.60 &  \textbf{57.51} & \textbf{51.54} \\
CIFAR10         &   31.72   &  38.93  & 36.40 &  40.36 & 41.60  \\
\bottomrule
\end{tabular}
\end{sc}
\end{small}
\end{center}
\end{table*}
\subsection{Noise Response}

\begin{figure}[t]
\centering 
\begin{subfigure}[t]{0.47\linewidth}
\includegraphics[width=1\textwidth]{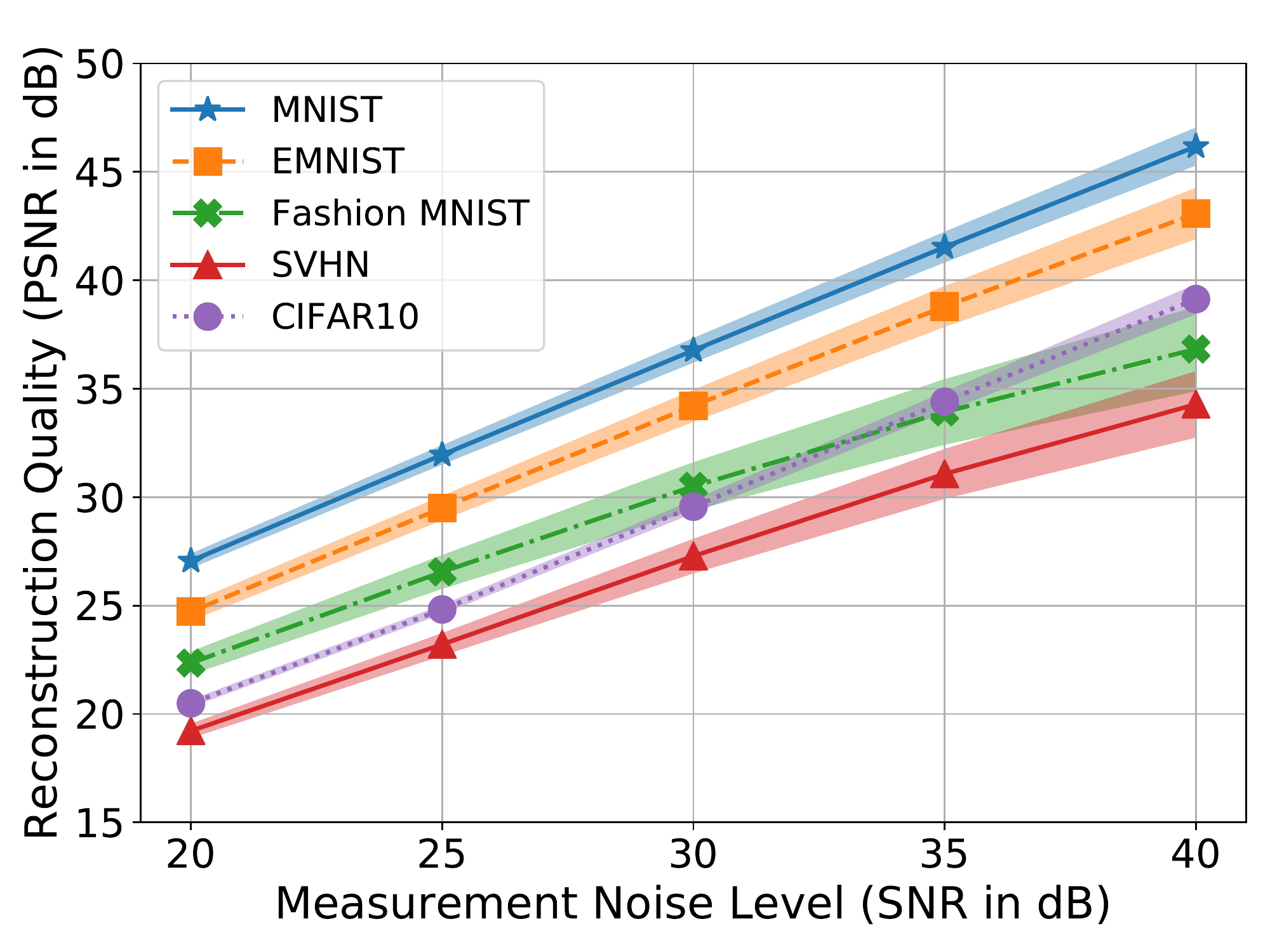}
\caption{Gaussian}
\end{subfigure}
~~
\begin{subfigure}[t]{0.47\linewidth}
\includegraphics[width=1\textwidth]{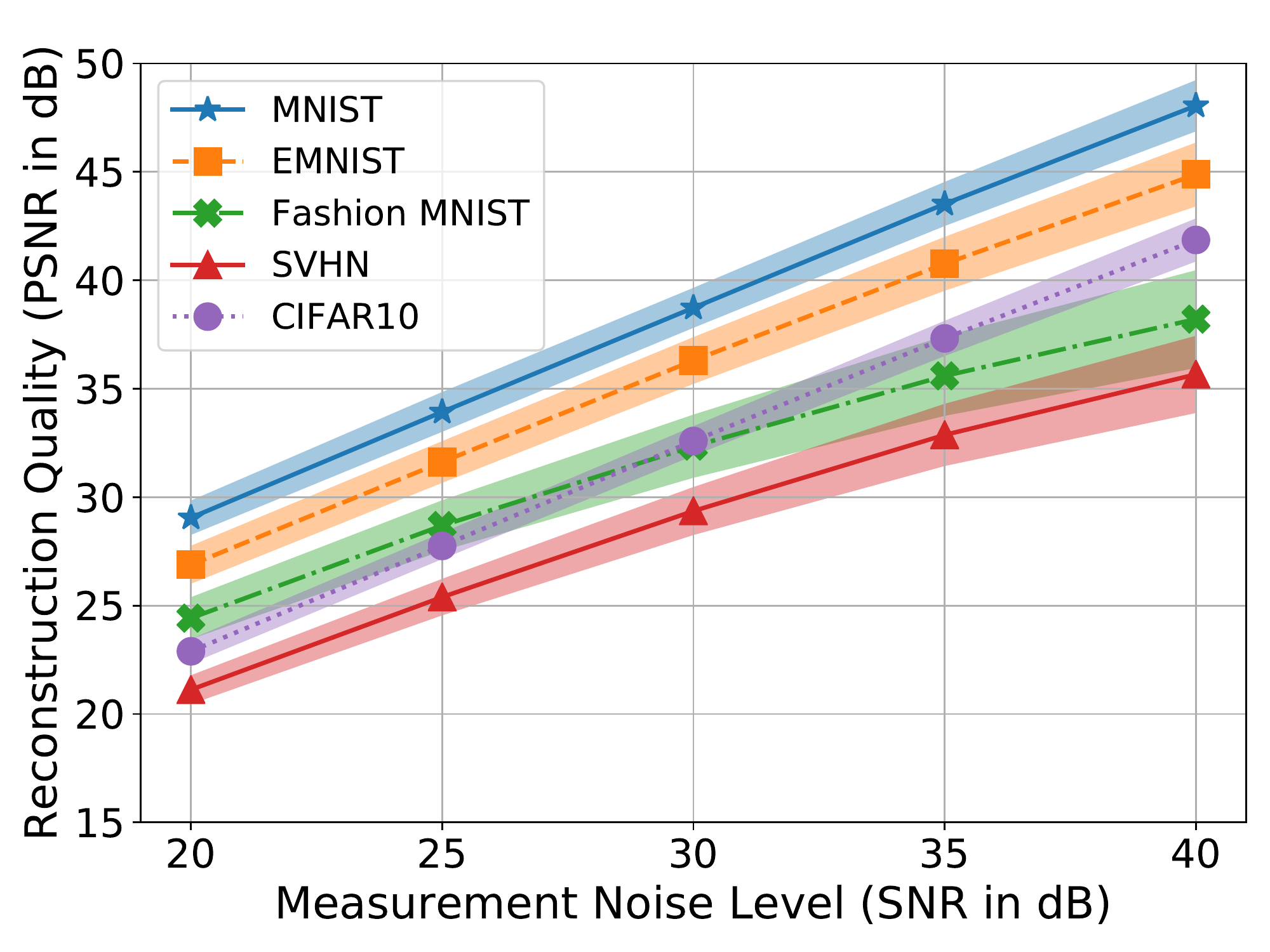}
\caption{Poisson}
\end{subfigure}
\caption{Reconstruction quality of the test images vs noise level of the measurements for different datasets. We learned the reference using noise-free measurements.}
\label{fig:noise}
\end{figure}

To test the robustness of our method in the presence of noise, we added Gaussian and Poisson noise at different levels to the measurements. Poisson noise or shot noise is the most common in the practical systems. We model the Poisson noise following the same approach as in \cite{metzler2018prdeep}. We simulate the measurements as 
\begin{equation}
    y(i)=|z(i)|+ \eta (i) \;\;\;\text{for all }  i=1,2,\ldots, m, 
\end{equation}
where $ \eta (i) \sim \mathcal{N}(0,\sigma^2)$ for Gaussian noise and $ \eta (i) \sim \mathcal{N}(0,\lambda|z(i)|)$ for Poisson noise with $z=Ax+Bu$. We varied $\sigma,\lambda$ to generate noise at different signal-to-noise ratios. Poisson noise affects the larger measurements with higher strength than the smaller measurements. As the sensors can measure only positive measurements, we kept the measurements positive by applying ReLU function after noise addition. 
% We expect the reconstruction to be affected by noise as we did not use any denoiser. 
We can observe the effect of noise in Fig.~\ref{fig:noise}. Even though we did not add noise during training, we get reasonable reconstruction and performance degrades gracefully with increased noise. 
% The relationship between noise level and reconstruction performance also indicates that our phase retrieval system is quite stable.

\subsection{Random Reference versus Learned Reference}
To demonstrate the advantage of the learned reference signal, we compared the performance of learned reference and random reference on some standard images. The results are shown in Fig.~\ref{fig_big_imgs}. The learned reference is trained using 32 images from CelebA dataset which we resized to $200\times200$. The test images used in Fig.~\ref{fig_big_imgs} are $512\times 512$, so we resized the learned reference from $200\times200$ to $512\times512$. For random reference, we selected the entries of the reference uniformly at random from $[0,1]$. We selected the best result out of 100 trials for every test image with random reference. We can observe from the results that our learned reference significantly outperforms the random reference even though the test image distribution is distinct from the training data. The number of steps of the unrolled network is $K = 50$.

\subsection{Comparison with Existing Phase Retrieval Methods}
We have shown comparison with other approaches in Table~\ref{table_different_methods}. We selected Kaczmarz \cite{wei2015solving} and Amplitude flow \cite{chen2015solving} for comparison using PhasePack package \cite{chandra2017phasepack}. We also show Hybrid Input Output (HIO), which is similar to our phase retrieval routine without any reference. We observe that our approach with learned reference can outperform all other approaches on all the datasets. All the traditional phase retrieval methods suffer from the trivial circular shift, rotation, and flip ambiguities, thus produce significantly worse reconstruction than our method does.
%
% prDeep \cite{metzler2018prdeep} is one of the deep learning based approaches which solves Fourier phase retrieval using a deep convolutional network as denoiser. However it performs 50 or so random initializations using HIO to select an initial signal that is provided to prDeep solver (select best out of 3 runs afterwards). To be fair to prDeep and other PR methods, they need multiple initializations because the Fourier PR problem is challenging without the additional reference. 
%
Our method uses a reference signal to simplify the initialization and removes the shift/reflect ambiguities. To mathematically explain this fact, a shifted or flipped version of $x$ would not give us the same Fourier measurements as $|F(x+u)|$ if $u$ is chosen appropriately as we do with the learning procedure. As we showed in Fig.~\ref{fig:noise}, our method can perfectly recover the shifted and flipped versions of the images using the  reference that was trained with upright and centered images. 
% In terms of run time, prDeep requires around 345 sec \cite{metzler2018prdeep} to recover a single image in Fig.~\ref{fig_big_imgs} at $256\times256$ size while our method takes only 0.41 sec. 
% Our method requires on average 0.41 seconds to recover a  $256\times256$ size. 

\begin{table*}[t]
\caption{Comparison with existing phase retrieval methods}
\label{table_different_methods}
\begin{center}
\begin{small}
\begin{sc}
\begin{tabular}{lcccccc}
\toprule
Methods & MNIST & EMNIST & F. MNIST  & SVHN & CIFAR10 \\
\midrule
HIO                     & 9.04  & 8.42    & 9.65    & 19.87 & 14.70  \\
Amplitude Flow          & 9.99  & 9.79     & 11.90   & 20.25 & 15.04  \\
Kaczmarz                & 11.81  & 11.47   & 13.44   & 19.48 &15.01  \\
Flat Reference          & 18.21 & 17.24   & 16.56   & 20.89 & 15.81 \\
Random Reference        & 36.87 & 28.41   & 27.27   & 36.45 & 25.57 \\
Learned Reference (Ours)& \textbf{66.54} & \textbf{58.72}   & \textbf{57.81}   & \textbf{57.51} & \textbf{41.60} \\
\bottomrule
\end{tabular}
\end{sc}
\end{small}
\end{center}
\end{table*}

\subsection{Effects of Number of Layers (K)}
\begin{figure}[t]
\centering 

\begin{subfigure}[b]{0.31\linewidth}
\centering
\includegraphics[width=1\linewidth]{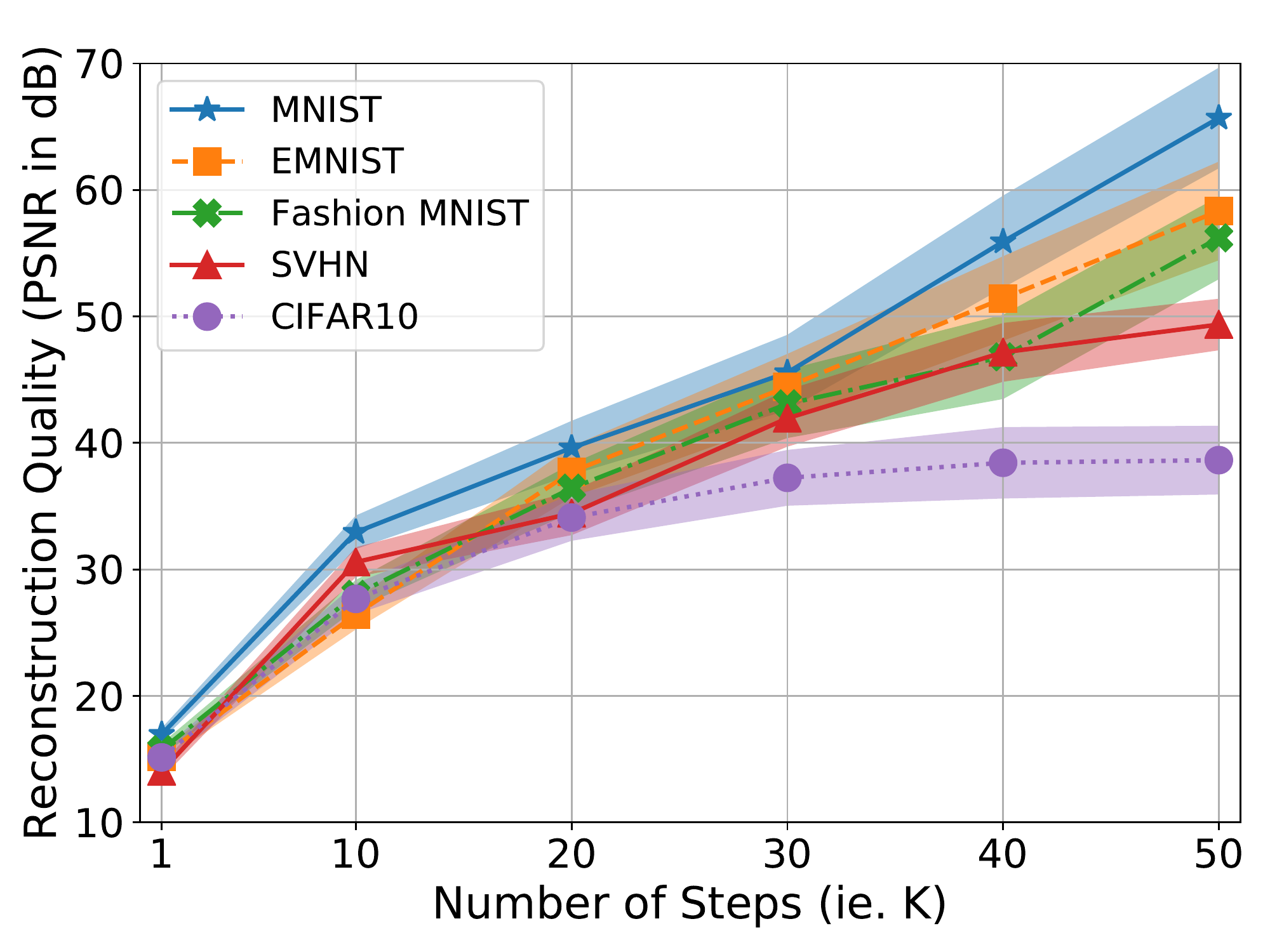}
\caption{Training K=Testing K}
\end{subfigure}
~
\begin{subfigure}[b]{0.31\linewidth}
\centering
\includegraphics[width=1\linewidth,trim={0 0 0 0},clip]{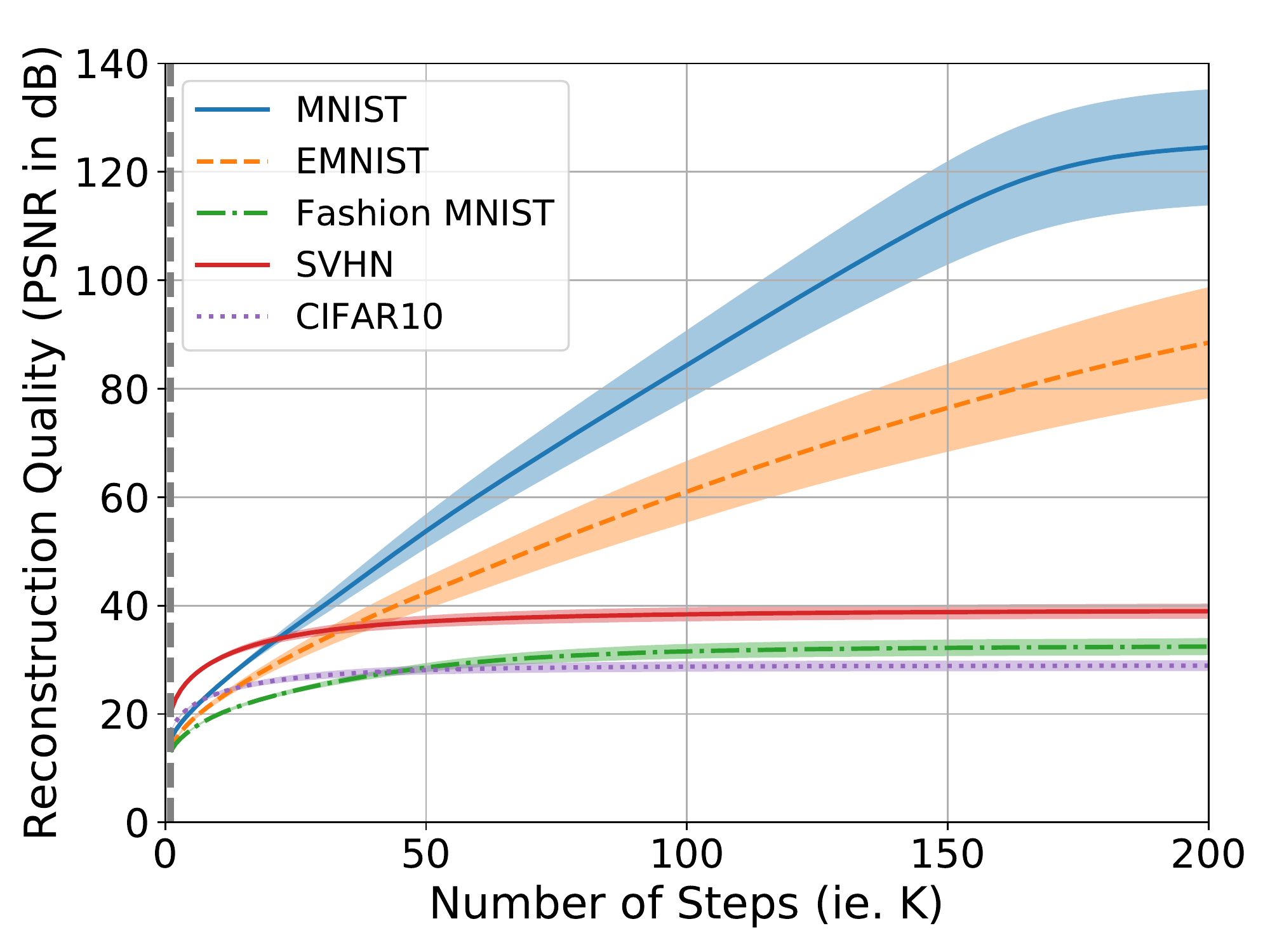}
\caption{Training K=1}
\end{subfigure}
~
\begin{subfigure}[b]{0.31\linewidth}
\centering
\includegraphics[width=1\linewidth,trim={0 0 0 0},clip]{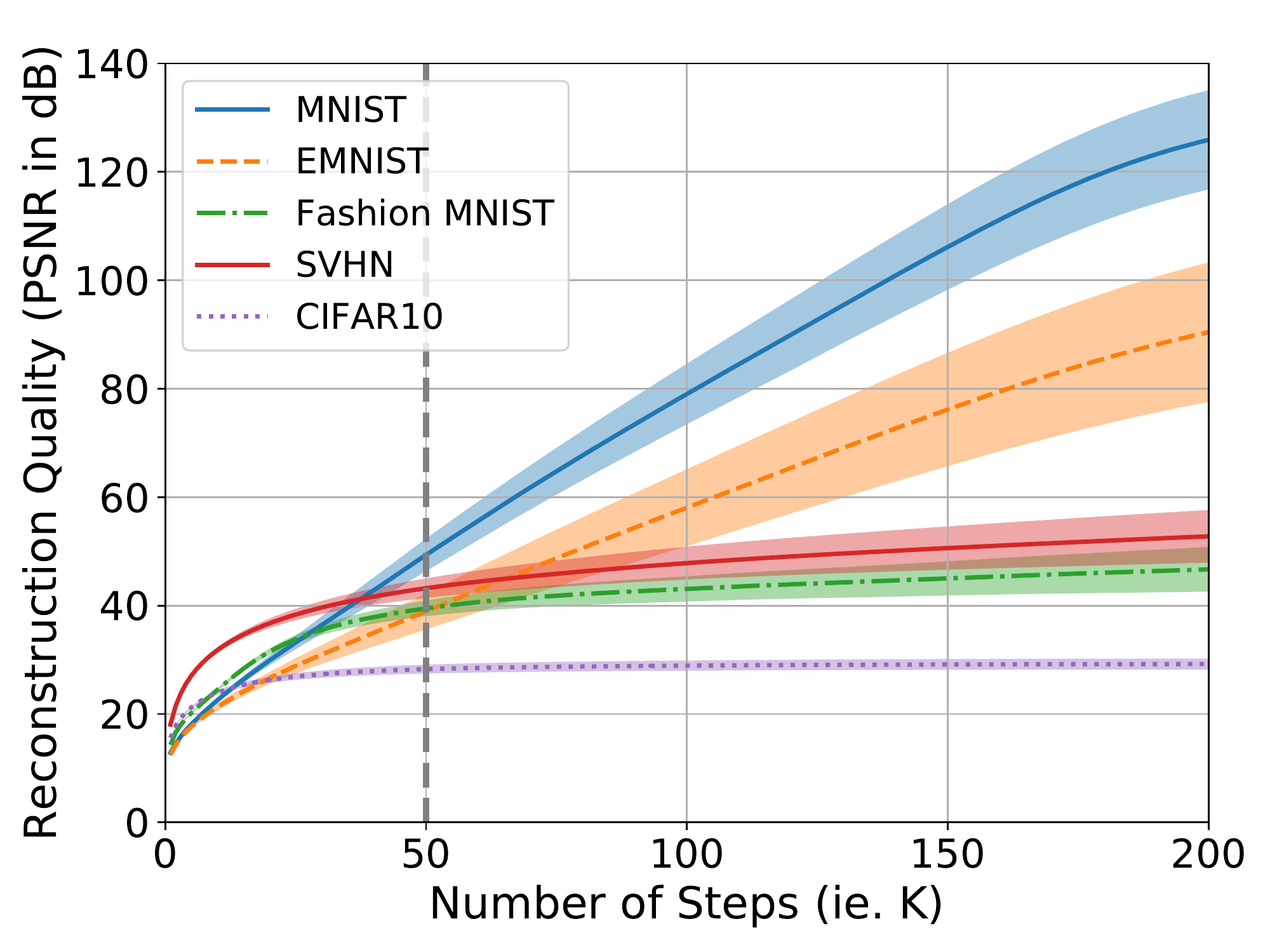}
\caption{Training K=50}
\end{subfigure}
\caption{Reconstruction PSNR vs the number of blocks ($K$) in the unrolled network at training and testing. (a) $K$ is same for training and testing (shaded region shows $\pm0.25$ times \texttt{std} of PSNR). (b) $K=1$ and (c) $K=50$, but tested using different $K$.}
\label{fig:diff_k}
\end{figure}

We tested our unrolled network with different numbers of layers (i.e., $K$) at training and test time. The results are summarized in Fig.~\ref{fig:diff_k}. 
We first used the same values of $K$ for training and testing. We observed that as $K$ increases, the reconstruction quality (measured in PSNR) improves. Then we fixed $K=1$ or $K=50$ at training, but used different values of $K$ at testing. We observed that if we increase $K$ at the test time, PSNR improves up to a certain level and then it plateaus. The PSNR achieved with reference trained with $K=50$ is better than what the referenced trained with $K=1$ provided. These results provide us a trade-off between the reconstruction speed and quality. As we increase $K$, the reconstruction quality improves but the reconstruction requires more steps (computations and time). 

\begin{figure}[t]
\begin{center}
\centerline{\includegraphics[width=\textwidth,trim={0cm 19cm 12cm 0cm},clip]{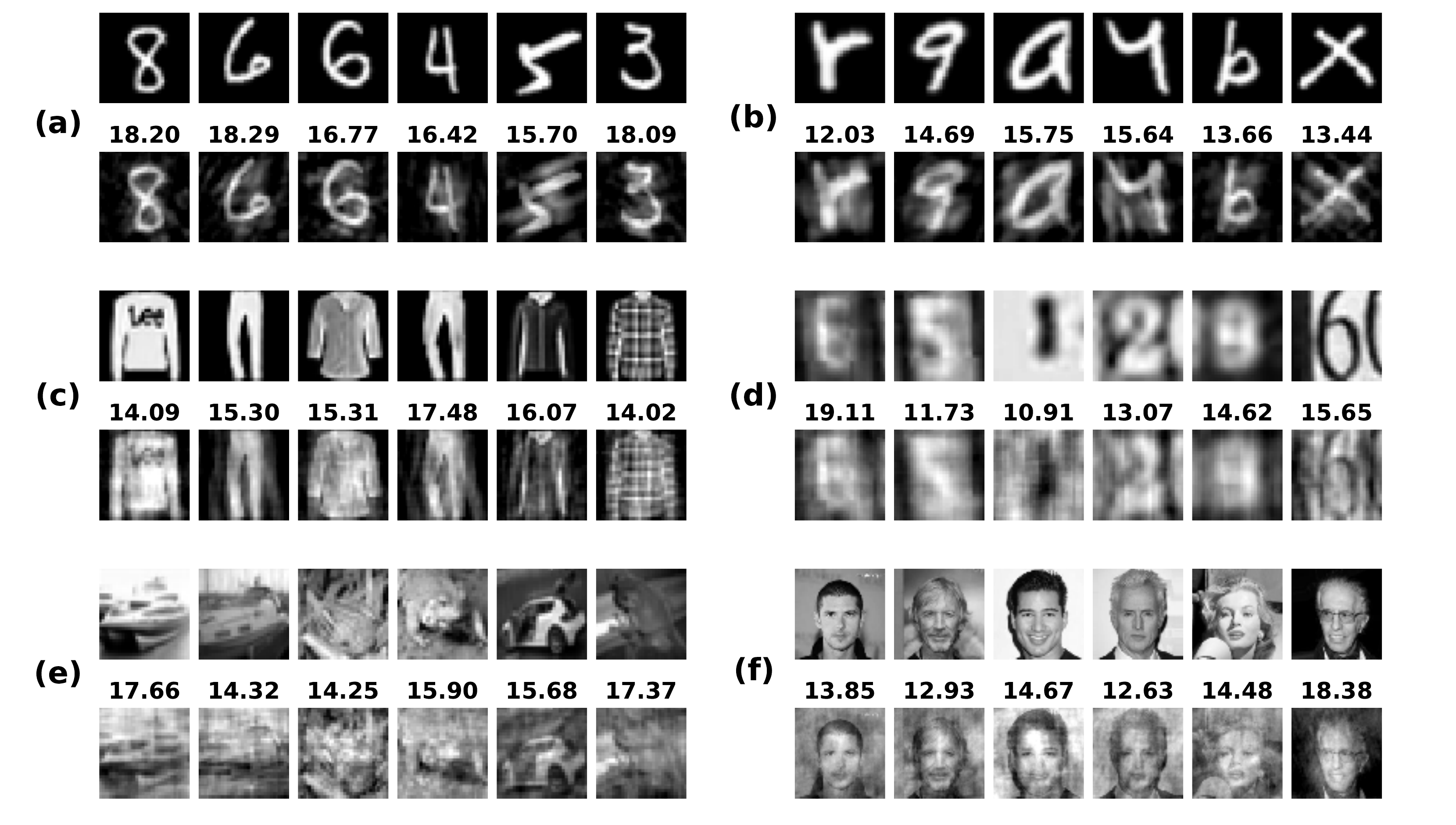}}
\caption{Single step reconstruction with reference in range $[0,1]$. Each of the \textbf{6 sets (a)-(f)} has the the ground truth in the first row. Second row is the reconstruction (PSNR shown on top of images.)}
\label{fig_1step}
\end{center}
\end{figure}

Finally, we learned a reference using $K=1$ and tested it on different images with $K=1$. To our surprise, our method was able to produce reasonable quality reconstruction with this extreme setting. We present some single-step reconstructions of each data set in Fig.~\ref{fig_1step}.

\subsection{Localizing the Reference}
\begin{figure}[!h]
\centering 
\begin{subfigure}[b]{0.47\linewidth}
\includegraphics[width=1\textwidth]{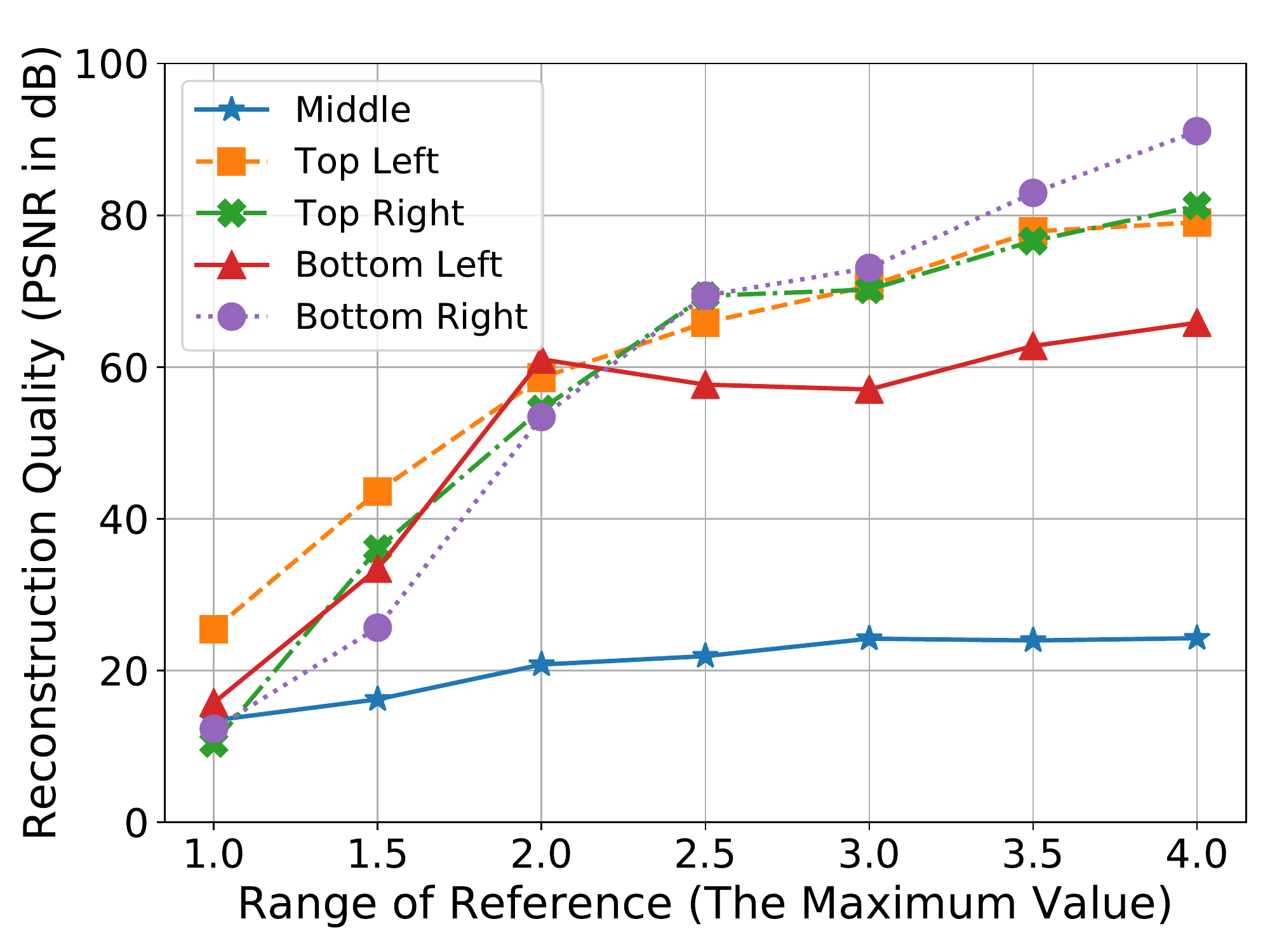}
\caption{MNIST}
\end{subfigure}
~~
\begin{subfigure}[b]{0.47\linewidth}
\includegraphics[width=1\textwidth]{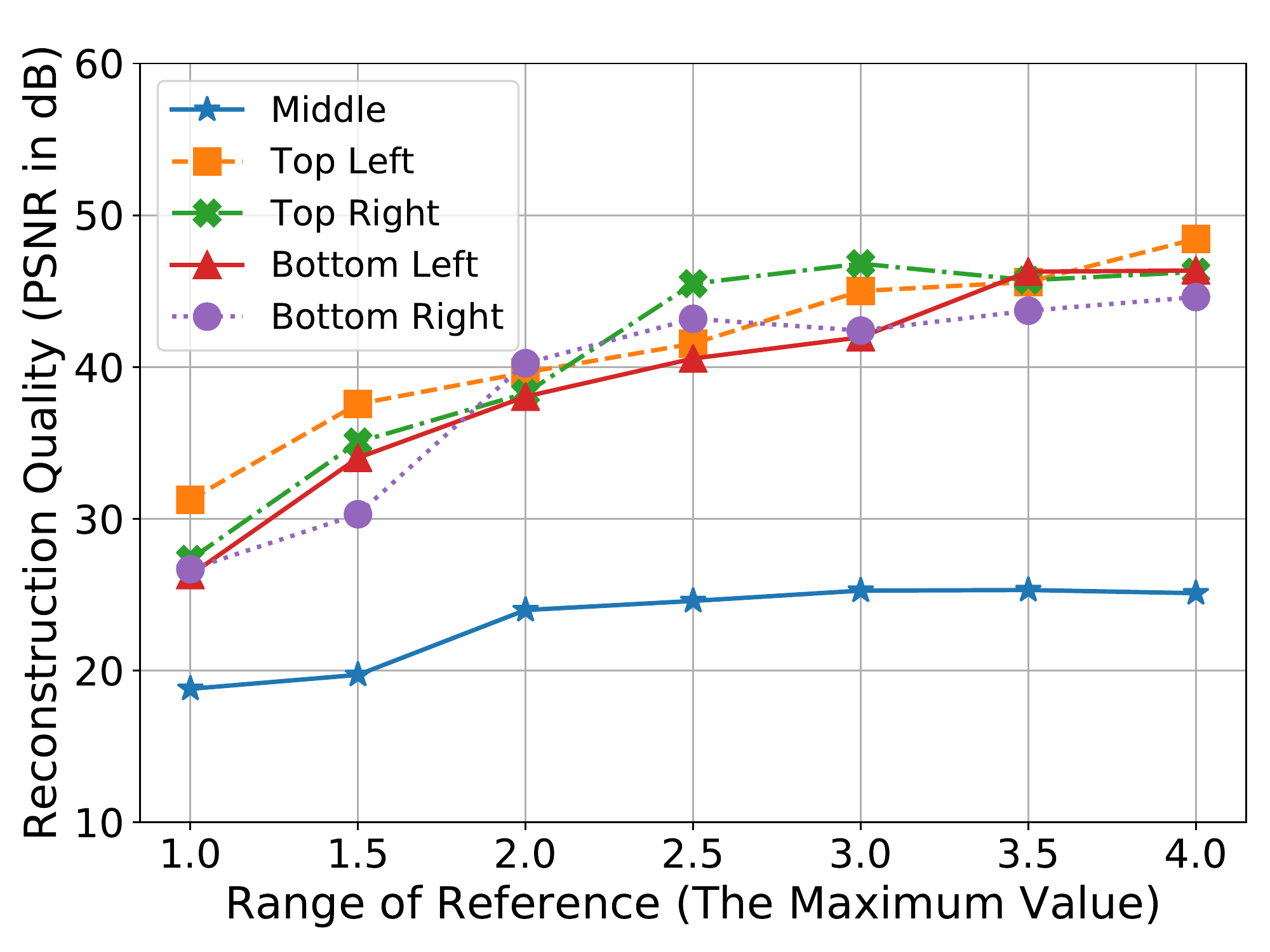}
\caption{CIFAR10}
\end{subfigure}

\caption{Performance of our method if the reference is an  $8\times8$ block placed at different positions. Fixing the minimum value at 0, we increased the maximum value of the reference we learn. We observe that the small reference placed in the corners performs better than the  ones placed in the center. }
\label{fig:masked_u}
\end{figure}

We also evaluated the effect of localizing the reference to a small region. For example, the reference is constrained to be within a small block in the corner or the center of the target signal. We restricted $u$ to be an $8\times 8$ block and placed it in different positions. We found that corner positions provide better results as shown in Fig.~\ref{fig:masked_u}. As we bring the reference support closer to the center, the quality of reconstruction deteriorates. 
This observation is related to the method in  \cite{barmherzig2019holographic,guizar2007holography,arab2020fourier}, where if the known reference signal is separated from the target signal, then the phase retrieval problem can be solved as a linear inverse problem. 

Note that signal recovery from Fourier phase retrieval is equivalent to signal recovery from its autocorrelation. We can write the autocorrelation of target plus reference signals as $(x+u)\star(x+u) = x\star x + u\star u + x\star u + u \star x$. The first term is a quadratic function of $x$, the second term is known, and the last two terms are linear functions of $x$. If the supports for $x$ and $u$ are sufficiently separated, then we can separate the last two linear terms from the first two quadratic terms and recover $x$ by solving a linear problem. However, if $x$ and $u$ have a significant overlap, then we need to solve a nonlinear inverse problem as we do in this paper.

\section{Conclusion}
We presented a framework for learning a reference signal to solve the Fourier phase retrieval problem. The reference signal is learned using a small number of training images using an unrolled network as a solver for the phase retrieval problem. Once learned, the reference signal serves as a prior which significantly improves the efficiency of the signal reconstruction in the phase retrieval process. The learned reference generalizes to a broad class of datasets with different distribution compared to the training samples. We demonstrated the robustness and efficiency of our method through extensive experiments.

\section*{Acknowledgments}
We would like to thank the anonymous reviewers for their insightful comments and suggestions. The first two authors contributed equally in this work. This research was supported in parts by an ONR grant N00014-19-1-2264, DARPA REVEAL Program, and a Google Faculty Award. 

% Note use of \abovespace and \belowspace to get reasonable spacing
% above and below tabular lines.

\bibliographystyle{splncs04}
\bibliography{refs}


\end{document}